\documentclass[12pt]{iopart}

\usepackage{graphicx}
\usepackage{bm}
\usepackage{subfigure} 

\begin{document}

\title{Factorization of correlations in two-dimensional percolation on the plane and torus}
\author{Robert M. Ziff$^1$, Jacob J. H. Simmons$^2$, and Peter Kleban$^3$ \\
$^1$MCTP 
and Department of Chemical Engineering, University of Michigan, 
Ann Arbor, MI 48109-2136 USA,
$^2$James Franck Institute, University of Chicago, Chicago, IL 60637 USA,
$^3$LASST and Department of Physics \& Astronomy,
University of Maine, Orono, ME 04469, USA
\eads{\mailto{rziff@umich.edu}, \mailto{simmonsj@uchicago.edu}, \mailto{kleban@maine.edu}} }

\begin{abstract}
Recently, Delfino and Viti have examined the factorization
of the three-point  density correlation function $P_3$ at the percolation point
in terms of the two-point density correlation functions $P_2$. According to conformal invariance, this factorization is exact
on the infinite plane, such that the ratio
$R(z_1, z_2, z_3) = P_3(z_1, z_2, z_3)/
[P_2(z_1, z_2) P_2(z_1, z_3) P_2(z_2, z_3)]^{1/2}$
is not only universal but also a constant, independent of the $z_i$, and in fact an operator product expansion (OPE) coefficient. Delfino and Viti analytically calculate its
value ($1.022013\ldots$) for percolation, in agreement 
with the numerical value 1.022 found previously in a study of $R$ on the conformally equivalent
cylinder.   In this paper we confirm the factorization on the plane numerically 
using periodic lattices (tori) of very large size, which locally approximate a plane.
We also investigate the general behavior of $R$ on the torus, and find a minimum value of $R \approx 1.0132$ when the three points are maximally separated. In addition, we present a simplified expression for $R$ on the plane as a function of the SLE parameter $\kappa$.
\end{abstract}

\maketitle

\section{Introduction}

The study of correlations in percolation provides insight into the  nature
of the percolation process.  The well-known two-point density correlation function
$P_2(z_1, z_2)$ as a function of the locations of the points $z_1$ and
$z_2$ behaves as
\begin{equation}
P_2(z_1, z_2) \sim |z_1 - z_2|^{2(D - d)}
\label{eq:P2}
\end{equation}
for large $|z_1 - z_2|$, on a $d-$dimensional percolating system at the critical point $p_c$, 
where $D$ is the fractal dimension, which has the universal value $91/48$ in two dimensions.
However, the coefficient to (\ref{eq:P2}) depends upon the model of percolation and also vanishes
in the continuum limit where the lattice spacing goes to zero, and is thus
non-universal.  

In order to study higher-order correlations, the present authors considered the ratio
\cite{KlebanSimmonsZiff06}
\begin{equation}
R(z_1, z_2, z_3) := \frac{P_3(z_1, z_2, z_3)}{\sqrt{P_2(z_1, z_2) P_2(z_1, z_3) P_2(z_2, z_3)}} \ ,
\label{eq:R}
\end{equation}
where $P_3(z_1, z_2, z_3)$ is the three-point density correlation function.  With the ratio defined this way (including the square root in 
the denominator), the lattice factors cancel
out and the quantity $R(z_1, z_2, z_3)$ converges to a universal function
in the continuum limit.
It was shown in \cite{KlebanSimmonsZiff06,SimmonsKlebanZiff07b} via conformal field theory
that if $z_1$ and $z_2$ reside on the boundary
of a (compact) bounded or half-infinite system  and $z_3$ is on the boundary or inside it,  then, in the continuum limit,
$R$ is a constant
independent of $z_1$, $z_2$, and $z_3$ and equal to
\begin{equation}
C_0 :=  2^{7/2} \pi^{5/2} 3^{-3/4} \Gamma(1/3)^{-9/2} = 1.0299268 \ldots \ .
\label{C0eq}
\end{equation}   This
behavior was termed factorization, i.e., the three-point function factors into a product of square roots of
two-point functions, multiplied by a constant.  In \cite{SimmonsZiffKleban09},
this concept was generalized to the case where correlations between intervals on the boundary of a rectangle
and a single point $z_1$ inside was studied.  There, the factorization is not exact, but depends
upon the distance from the bounding intervals and the boundary conditions (free or wired--a wired interval means that all sites are constrained to belong to one cluster).
Far from the bounding intervals, $R$ once again approaches $C_0$.   
Related recent work includes Refs.\
\cite{HonglerSmirnov10,KarsaiKovacsAnglesdAuriacIgloi08,Ridout10,MathieuRidout07,MathieuRidout08,Ridout09,SimmonsCardy09,SheffieldWilson10}

Recently, Delfino and Viti \cite{DelfinoViti10} have examined the factorization  for three points on an infinite
plane for the general Potts model.  Here the factorization is exact, following simply from the general form of the three-point function with all three operators the same \cite{Polyakov70}.
However, it is not possible to find a general expression for $R$ (the OPE or operator product expansion coefficient) using methods specific to minimal CFTs
(conformal field theories), such as the result in \cite{Dotsenko85}.  In various models, difficulties may occur for a variety of reasons: operators with non-integer Kac indices, coefficients that vanish due to additional symmetries (i.e., in the Ising model, spin reversal symmetry means that $\langle \sigma \sigma \sigma \rangle=0$ regardless of cluster properties), or multiple fields with a common weight. By coupling the CFTs to Liouville gravity (LG), Al.~Zamolodchikov obtained OPE coefficient expressions \cite{AlZamolodchikov2005} that resolve the issues of non-integer Kac indices (as occurs, e.g., for percolation) and additional symmetries. 

In \cite{DelfinoViti10} Delfino and Viti used the LG result to find $R$. The  value they obtain is not identical to the LG three-point coefficient, because the LG analysis assumes a unique operator with each weight, but Delfino and Viti argue that the local selection rules of the Potts disorder operator $\mu_{\alpha \beta}$ are implemented by two identical weight fields $\mu$ and $\bar \mu$.  They then suggest that the LG analysis might still apply to a symmetric combination of these fields, which translates to an extra factor of $\sqrt{2}$ for the degenerate fields.  This gives $R = 1.0220\ldots =: C_1$.  In the Appendix, 
we show that the formula for $C_1$ of Delfino and Viti may be reduced to a single integral expression, and list numerical values for percolation as well as for other examples of the Potts model with integer $q$ values, including both low density (FK cluster) and high density (spin cluster) phases.  For percolation we indeed find
\begin{equation}
C_1 = 1.022013133\ldots
\end{equation}

The value $C_1 \approx 1.022$ was originally found numerically by the present authors when studying correlations between the two ends of a cylinder and 
a point $z_1$ in the interior \cite{SimmonsZiffKleban09}.  When two ends of the cylinder are ``wired," we also found numerically that $R$ approaches $C_1$ exponentially as $\exp(-2 \pi x/L)$
where $L$ is the dimension (circumference) of the end of the cylinder and $x$ is the distance from the nearest
end to $z_1$.  The correspondence between
the cylindrical results and the planar problem follows from the fact that the cylinder can be conformally mapped to the surface of
a sphere, so that the cylinder boundaries map onto two circles of equal radius.  In the limit of the infinitely long cylinder the radii of the image circles on the sphere shrink to zero; then the problem becomes that of the correlation of three points, and the factorization is everywhere exact, as 
 the sphere is conformally equivalent to the plane.

\begin{figure}
\centering
  \qquad \begin{minipage}{1.2in}t
\includegraphics[scale=0.4]{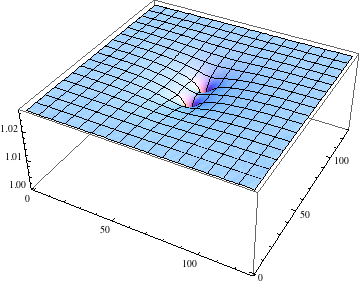}
  \includegraphics[scale=0.4]{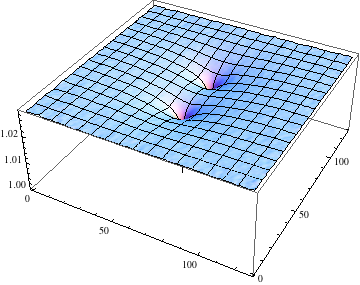}
   \includegraphics[scale=0.4]{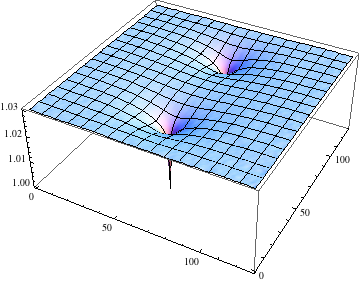}
    \includegraphics[scale=0.4]{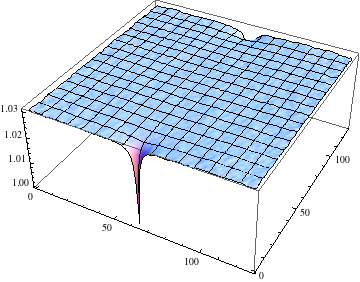}
  \end{minipage}  \normalsize
   \caption{(Color online) $R(z_1,z_2,z_3)$ on a system with open b.c.\ of $128\times128$ sites,
   as a function of $z_3$, for $z_1$ and $z_2$ fixed and 
   separated by $\Delta = |z_1 - z_2|$ given by (a) $16$, (b) $32$, (c) $64$, and (d) $128$  [top to bottom].
   At the boundaries, $R$ is approximately equal to (a) $1.0265$, (b) $1.0285$, (c) $1.0295$, and (d) $1.030$.
   }
   \label{fig:open}%
\end{figure}

\begin{figure}
\centering
\includegraphics[scale=0.5]{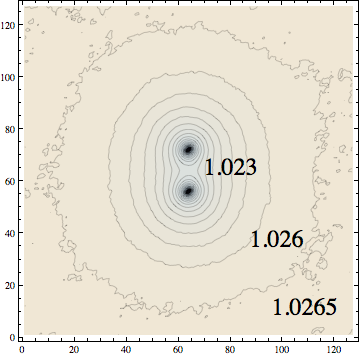}
\caption{\label{fig:opencontours}(Color online) 
Contours of $R$ of Fig.\ \ref{fig:open}(a) (open b.c.) with $\Delta = 16$ and $L = 128$.
The first complete contour encircling both points $z_1$ and $z_2$ is for $R = 1.023$, and $R$ increases
by $0.0005$ in each contour going outward. When $z_3$ goes to $z_1$ or $z_2$, 
$R \to 1$.}
\label{opencontoursfig}
\end{figure}

\begin{figure}
\centering
\includegraphics[scale=0.5]{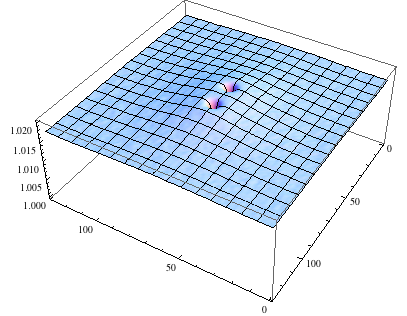}
\caption{\label{fig:Rperiodic3d}(Color online) 
$R(z_1, z_2, z_3)$ for a system with periodic b.c.\ of $128 \times 128$ sites, with
the two 
points $z_1$ and $z_2$ separated by a distance 
$\Delta = 16$, as a function of the third point $z_3$.
Near $z_1$ and $z_2$,  $R$ rises to a maximum value of about $1.0205$,
and drops to a value of about $1.018$ far from those points (at the edges
in this representation of the torus).  Contours are shown in Fig.\ \ref{fig:contoursperiodic}.}
\end{figure}

\begin{figure}
\centering
\includegraphics[scale=0.5]{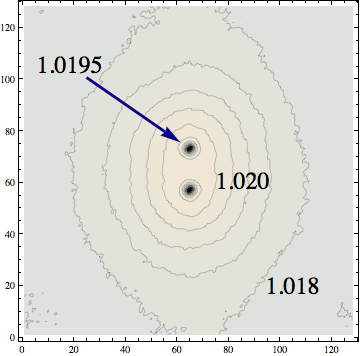}
\caption{\label{fig:contoursperiodic}(Color online) 
Contours of $R$ for the periodic system (torus) of Fig.\ \ref{fig:Rperiodic3d}
with $\Delta = 16$ and $L = 128$.
The first complete contour encircling both fixed points near the center is at $R = 1.020$, and $R$ decreases
by $0.0005$ in each contour going both inward and outward from that contour.}
\end{figure}

\begin{figure}
\centering
\includegraphics[scale=0.5]{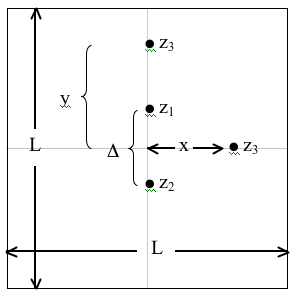}
\caption{\label{fig:PointFig}
Location of the fixed points $z_1$ and $z_2$, and the variable
point $z_3$ in the horizontal and vertical directions, assuming
$(x_0, y_0)$ is centered at the origin.  }
\end{figure}

In this paper, we consider the problem of measuring
the three-point correlations on the
plane and on the torus.  For open boundary conditions, it is not 
possible to simulate a  system large enough to effectively probe the 
infinite-plane behavior.  By using a large periodic system  and taking
advantage of its translational symmetry,
we are able to see the factorization over length scales large compared
to the lattice spacing but small compared to the size of the system.
We also find interesting behavior of the correlations on the torus itself when 
points are separated by distances on the order of the size of the torus.

\section{Simulation method and results}

For most of our simulations, we consider bond percolation on square lattices of size $L \times L$ at the critical point $p_c = 1/2$.  The number of samples ranged from ${\cal O}(10^5)$ for the
largest system, to over $10^9$ for the smaller ones.  We carried out simulations with both open and periodic b.c.

\subsection{Open boundary conditions}
First we consider open boundary conditions on the square, for relatively small $L$.
We take $z_1$ and $z_2$ fixed about the center of the lattice and
separated by $|z_1-z_2| = \Delta$, and determine $R$ as a function of $z_3$, where $z_3$ can be anywhere on the plane.   The
simulation technique here is to grow one critical cluster from $z_1$, and add 1 to the value of an array $N_{13}(z_3)$
to every point $z_3$ that the cluster wets.   If the cluster reaches $z_2$, then all the wetted-site coordinates $z_3$ of 
the cluster are
also added to the arrays $N_{23}(z_3)$ and $N_{123}(z_3)$, and to the counter $N_{12}$ which tells if points 1
and 2 connect.  If the cluster does not reach $z_2$, then a new cluster is grown from
$z_2$ and all of its sites $z_3$ are added to the array $N_{23}(z_3)$.  Finally, we normalize all these quantities
by the number of runs to get the probabilities, and calculate $R(z_1, z_2, z_3)$ according to (\ref{eq:R}).  The results are shown in Fig.\ \ref{fig:open}
for four values of $\Delta$ and $L = 128$.

In all cases there is a downward-pointing spike $R \to 1$  around $z_1$ and $z_2$, as $R = 1$ is the exact value when
two points coincide.   Note the highly expanded
scale in these plots.  When $\Delta = 128$ [Fig.\ \ref{fig:open}(d)], the two points $z_1$ and $z_2$ are at the edge, and the
results of \cite{KlebanSimmonsZiff06} apply, so $R$ has the value $C_0 \approx 1.0299\ldots$ of equation (\ref{C0eq}) at every point
in space except for the spikes.
The size of the spikes (which decay as a power-law as the separation between $z_3$ and $z_1$ or $z_2$ increases) is controlled by the discreteness of the lattice, and can be understood theoretically \cite{SimmonsZiffKleban09}.

As $\Delta$ decreases, two things happen: the roughly constant value at the edge decreases (values are given in the caption to Fig.\ \ref{fig:open})
and also $R$ varies markedly over the whole lattice.  While for $z_3$ near the center $R$  has value $\approx 1.022$ predicted by
\cite{DelfinoViti10} (see Fig.\ \ref{opencontoursfig}), there is no extended region nearby where $R$ is constant.

We have looked at  larger lattices (up to $L = 512$) and find the size of the constant region near the two fixed points increases somewhat, suggesting much larger
lattices are needed to observe the infinite-plane behavior.  However, the poor statistics of this type of simulation (only one data point for a given triangle $z_1, z_2, z_3$ is found from each sample) makes going to larger lattices impractical.  To overcome this problem, we consider lattices with periodic b.c.\ (tori).

\subsection{Periodic boundary conditions}

With periodic boundary conditions, every point is equivalent by translational invariance, so it is possible to get $L^2$ data points on an $L \times L$ lattice
for a given triangle of points $z_1$, $z_2$ and $z_3$, and results in much 
better statistics.
However, the question of what effect the toroidal geometry imposed by these boundary conditions has, and how to extract the planar result that we are interested in remains.
We expect that for a large enough torus, the behavior for the three points separated by distances much less than $L$ should be the same as for a plane.  However, because
the density of correlations drops off very slowly  according to (\ref{eq:P2}), the influence of the periodic b.c.\ should remain strong across relatively large systems.

To contrast what happens with periodic vs.\ free b.c., we first consider a simulation similar to that done for the open b.c.\ system above, in which $z_1$ and $z_2$ are fixed with $\Delta = 16$, and $R$ is determined for all $z_3$
(thus not making use of the translational invariance).  Here (see Figs.\ 3 and 4) we find an interesting result: $R$ has a maximum of about $1.02$ near the center, but then for large distances drops to $\approx 1.018$, which is below the value that would be found on the cylinder or any surface transformed from it.
Presumably, this decrease is due to the effects of the periodic b.c.\ on $P_2$ and $P_3$.

For the rest of our simulations on periodic systems, we make use of the translational invariance by populating the entire lattice, and looking at specific configurations of the three points.
We consider every possible location of the two fixed points $z_1$ and $z_2$ arranged vertically and separated by a distance $\Delta$, and vary $z_3$ both in the horizontal direction (along the perpendicular to the centerline) and in the vertical direction, as shown
in Fig.\ \ref{fig:PointFig}.   Specifically, for each point $(x_0, y_0)$, we set $z_1 = (x_0, y_0+\Delta/2)$, $z_2 = (x_0, y_0-\Delta/2)$.  To vary in the horizontal direction we considered $z_3 = (x_0 + x, y_0)$ for a range in values of $x$, and for  the vertical direction, we considered $z_3 = (x_0, y_0 + y)$ for a range in values of $y$.  In the vertical case, we consider both $|y| < \Delta/2$, i.e., $z_3$ between the two points, and $|y| > \Delta/2$, outside the two points.

 In these simulations, we create clusters on the entire lattice 
using the growth algorithm, labeling each cluster with a different index, and then check the indices of the three
points in order to calculate $R$.  If a pair $i,j$ of the points belong to the same cluster, then we increment
an array $N_{ij}^{(h)} (x)$ or $N_{ij}^{(v)}  (y)$ by 1.
If all three points belong to the same clusters, we increment $N_{123}^{(h)} (x)$ or $N_{123}^{(v)} (y)$ by 1.   We considered $L = 128, 256, \ldots
16384$, the latter being the largest size that could easily be simulated in our computer.

\begin{figure}
\centering
  \qquad \begin{minipage}{3 in}
  \includegraphics[scale=0.3]{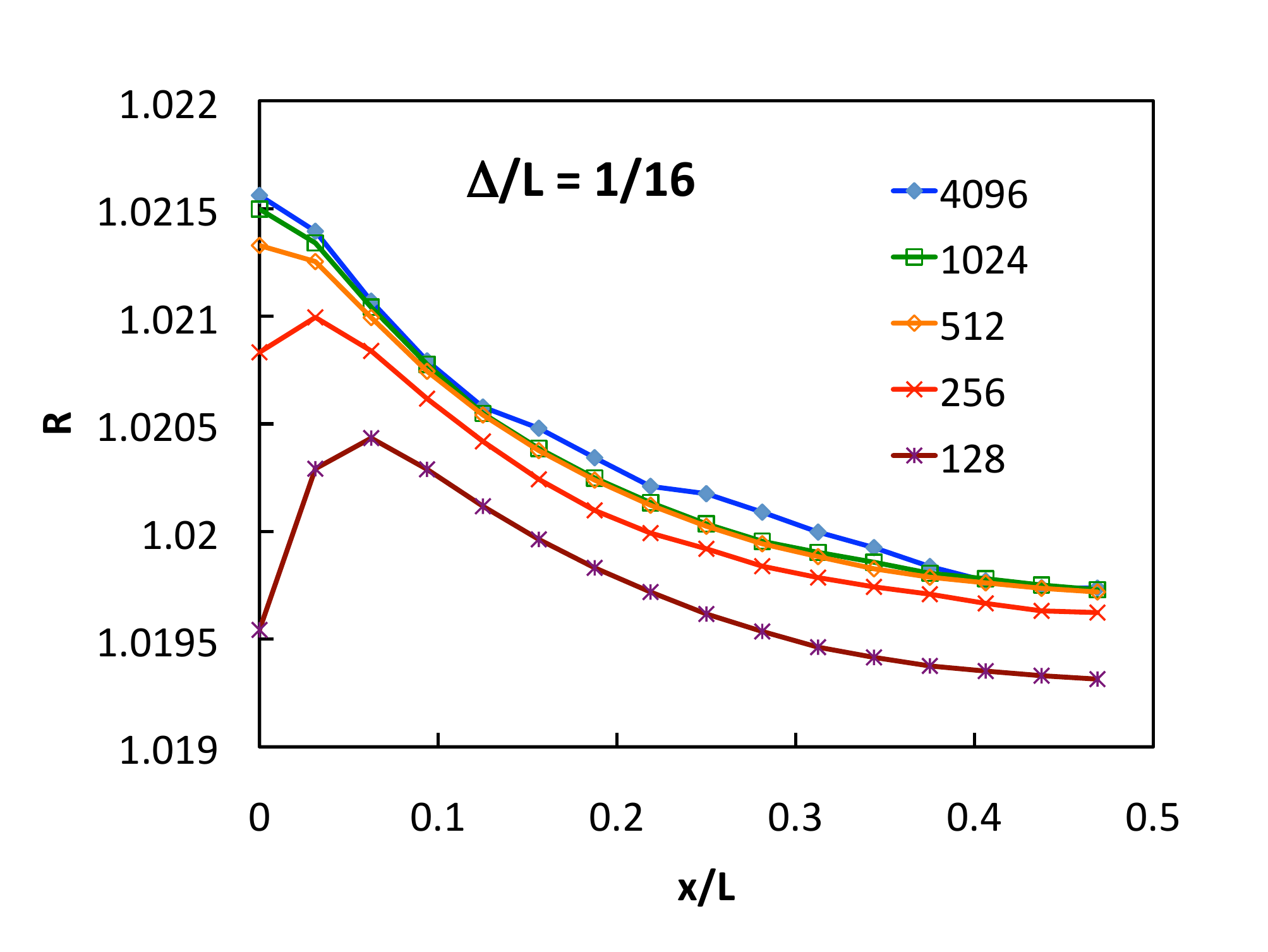}
\includegraphics[scale=0.3]{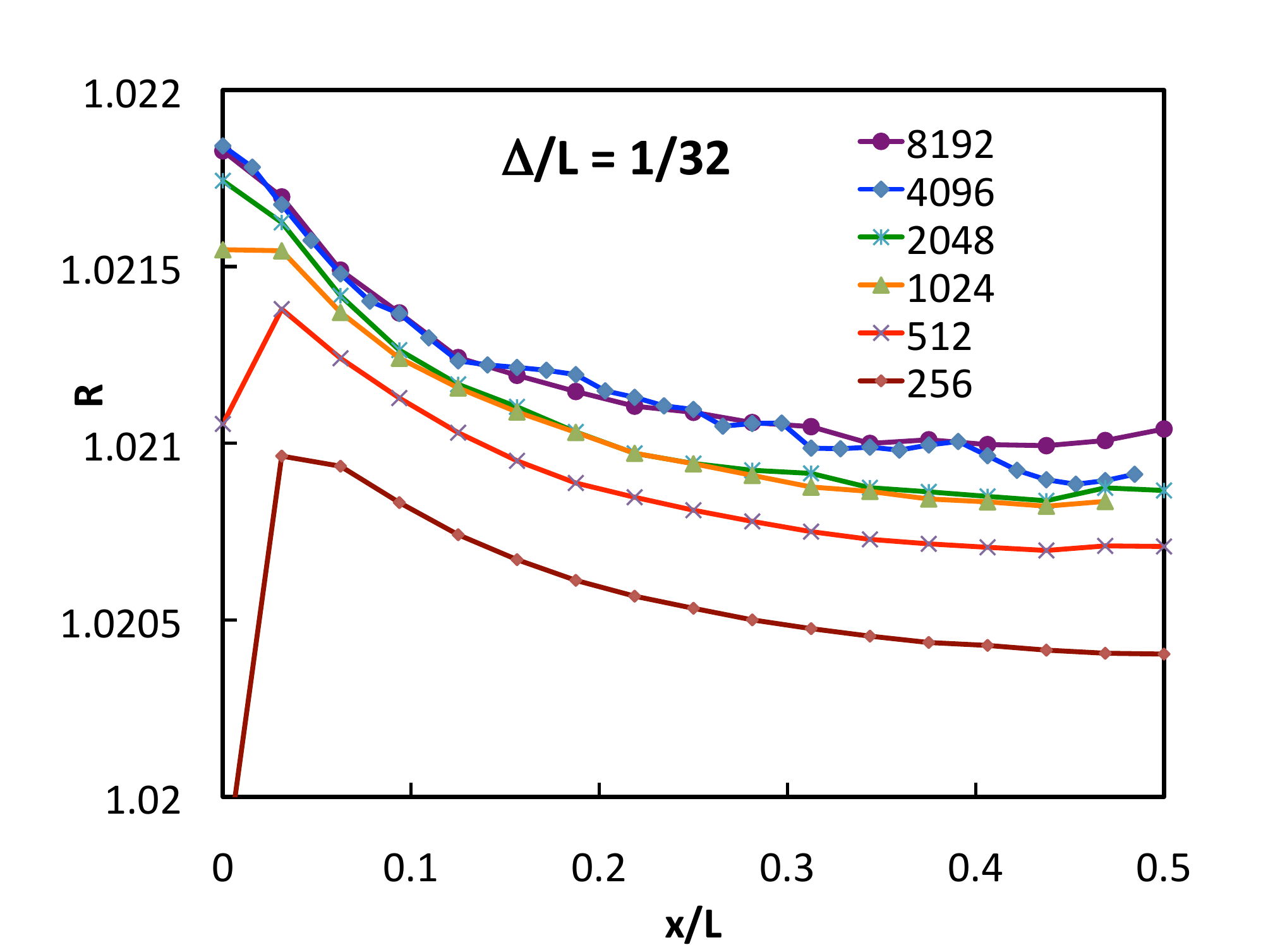}
  \includegraphics[scale=0.3]{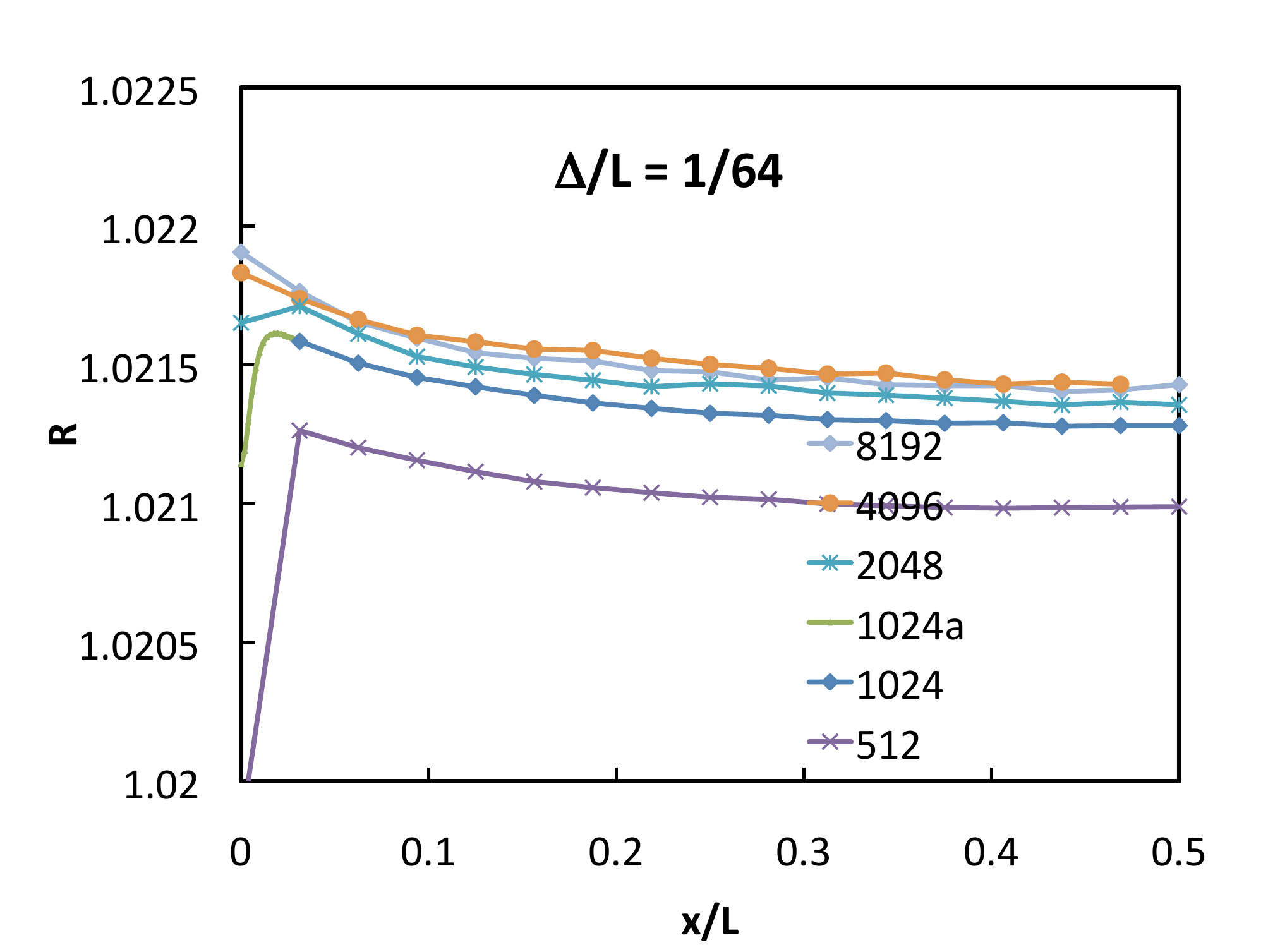}
   \includegraphics[scale=0.3]{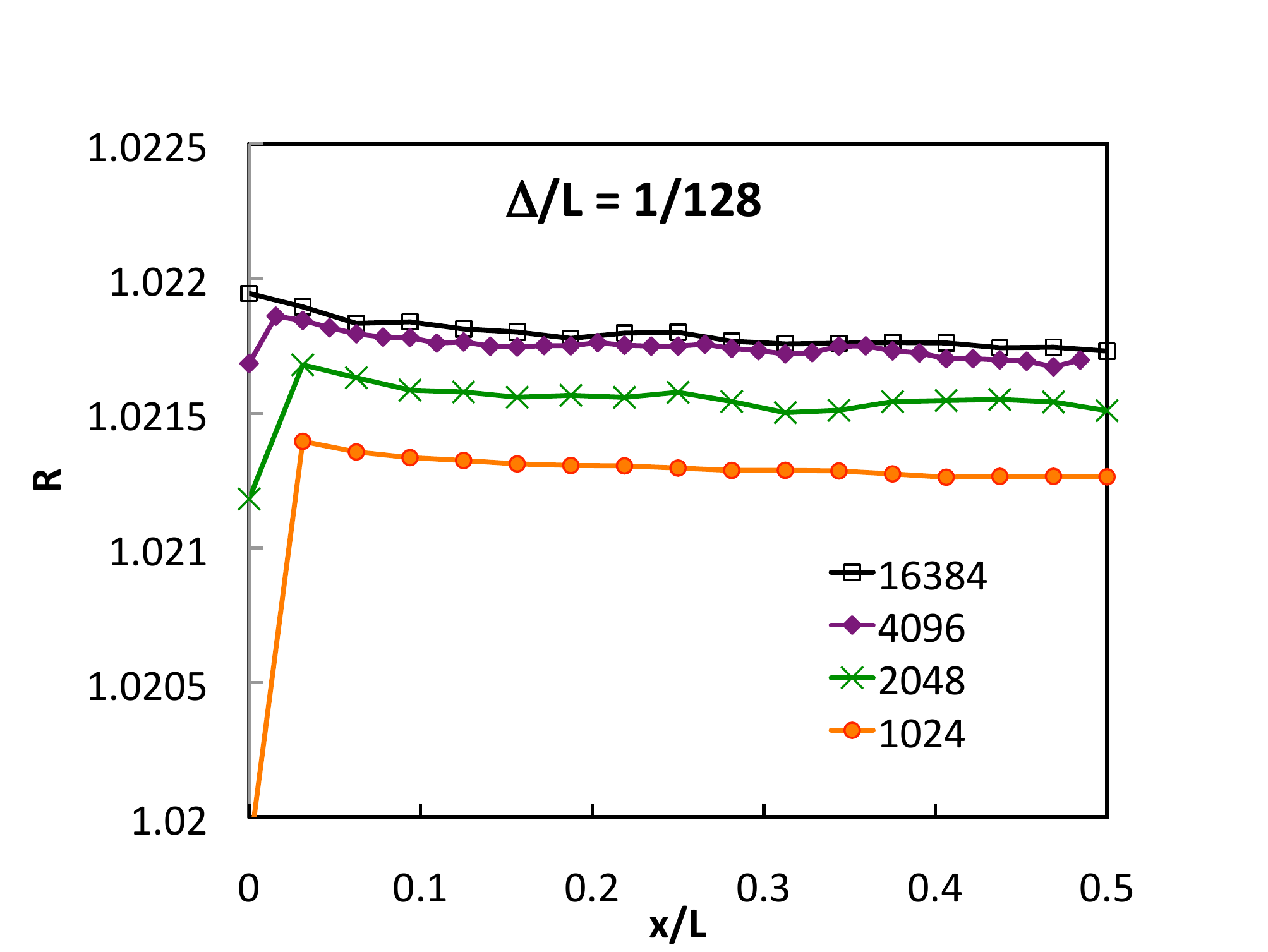}
    \includegraphics[scale=0.3]{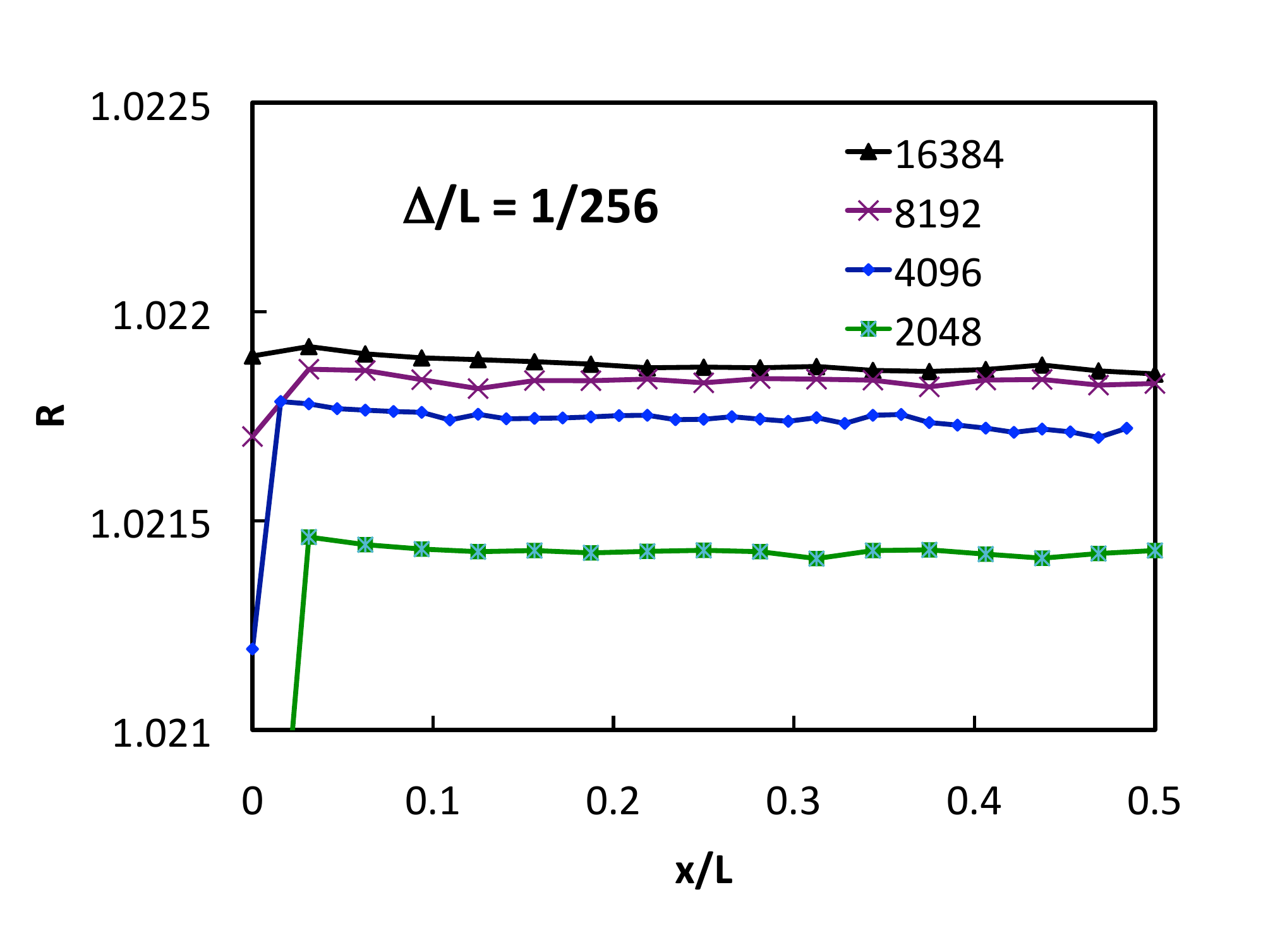}
  \end{minipage}
  \caption{\label{fig: Rperiodic}(Color online) 
$R$ as a function of $x/L$ on an $L \times L$ torus, with $\Delta/L = 1/16$, $1/32$, $1/64$, $1/128$, and $1/256$ (top to bottom), where $x$ is the distance from the center in the horizontal direction
for fixed ratios of $\Delta/L$ and various values of $L$ as given in the legends.  Note the different vertical scales.}
   \label{fig:Rperiodic}%
\end{figure}

Fig.\ \ref{fig:Rperiodic} shows the behavior of $R$ in the horizontal direction for a series of systems keeping $\Delta/L = 1/16 \ldots 1/256$ constant.   From these results we see:
\begin{itemize}
\item For small $\Delta$  ($\le 64$), there is a decrease in $R$ as $x \to 0$,  as the three points are within the distances in which finite-size lattice effects are significiant.
\item For large $x/L$ (and all $\Delta/L$), $R$ decreases as $x/L$ increases.  However, the amount of decrease becomes less as $\Delta/L$ decreases, and the curve of $R$ vs.\ $x/L$ becomes nearly
horizontal for the smallest $\Delta/L$ we consider ($1/256$).
\item As $L$ increases (for a fixed value of $\Delta/L$ and not too small $x/L$), the curves of $R$ as a function of $x/L$ approach a limiting form.
In fact, except for small $x/L$, the effect of increasing $L$ is simply to move the curves vertically.  This can be
understood as being due to finite-size effects on $P_{2}(z_1,z_2)$.
\item In particular, in the case $\Delta/L = 1/256$, for large $L$, $R$ is nearly independent of $L$ and is close to the expected value $R \approx 1.022$.
\end{itemize}

\begin{figure}
\centering
\includegraphics[scale=0.5]{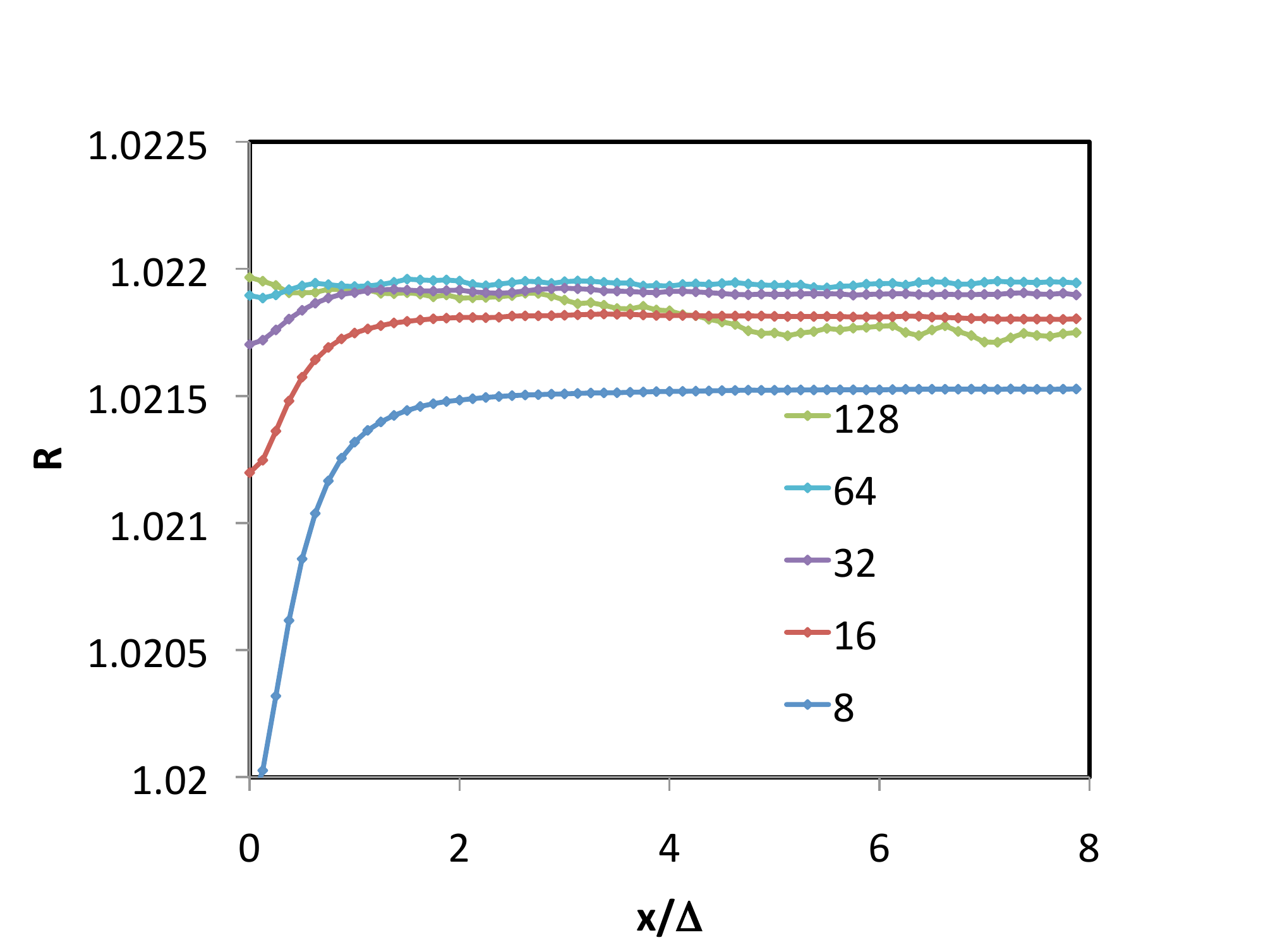}
\caption{\label{fig:Rclose}(Color online) 
Values of $R$ as a function of the scaled distance $x/\Delta$ to the point $z_3$ from the center of the pair of fixed points
in the horizontal direction, for values of $\Delta$ given in the legend, on a system with $L = 16384$. }
\end{figure}

\begin{figure}
\centering
\includegraphics[scale=0.5]{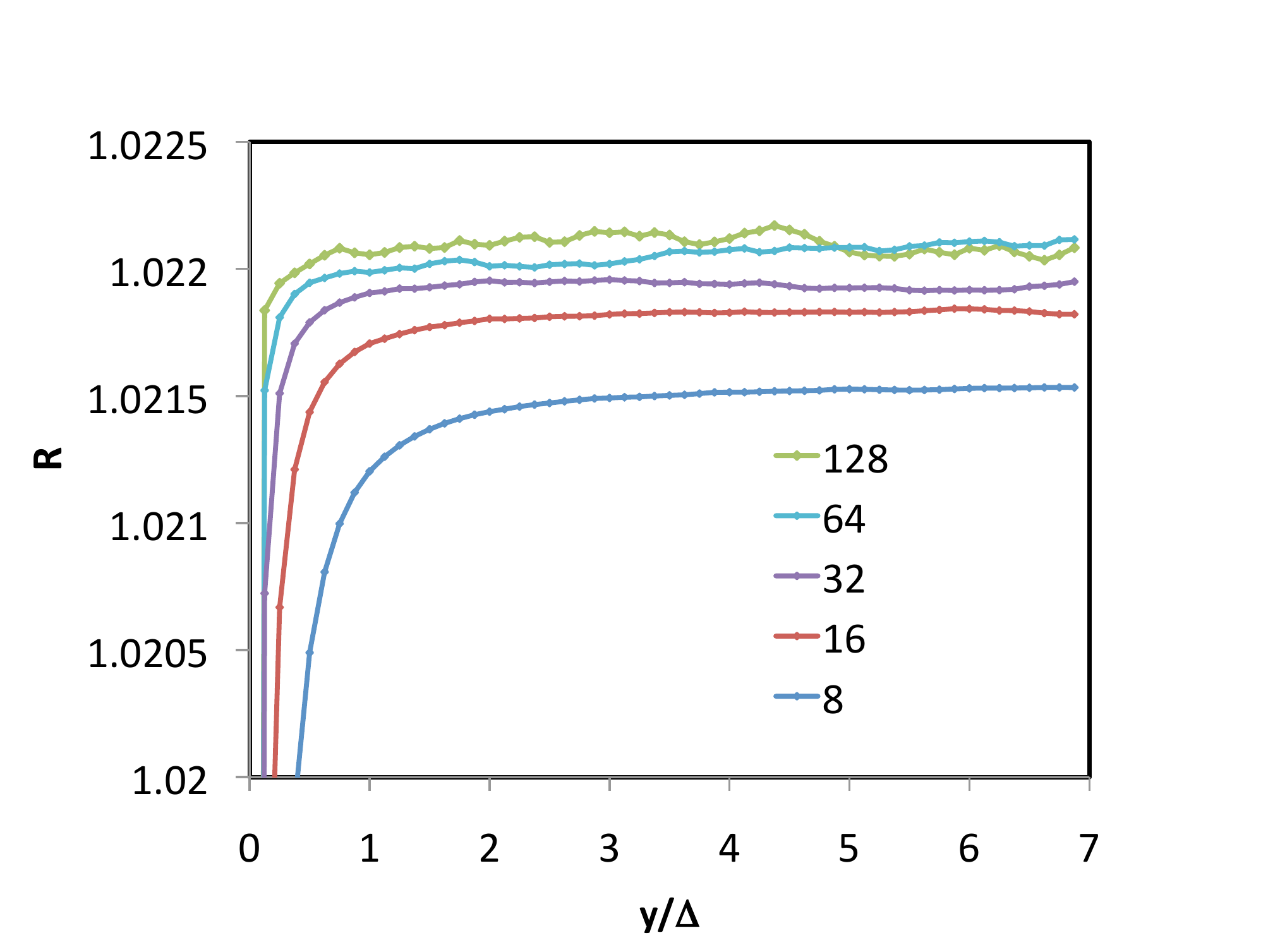}
\caption{\label{fig:Rvert}(Color online) 
Similar to Fig.\ \ref{fig:Rclose}, but in the vertical direction.  Legend values correspond to $\Delta$. }
\end{figure}

\begin{figure}
\centering
\includegraphics[scale=0.5]{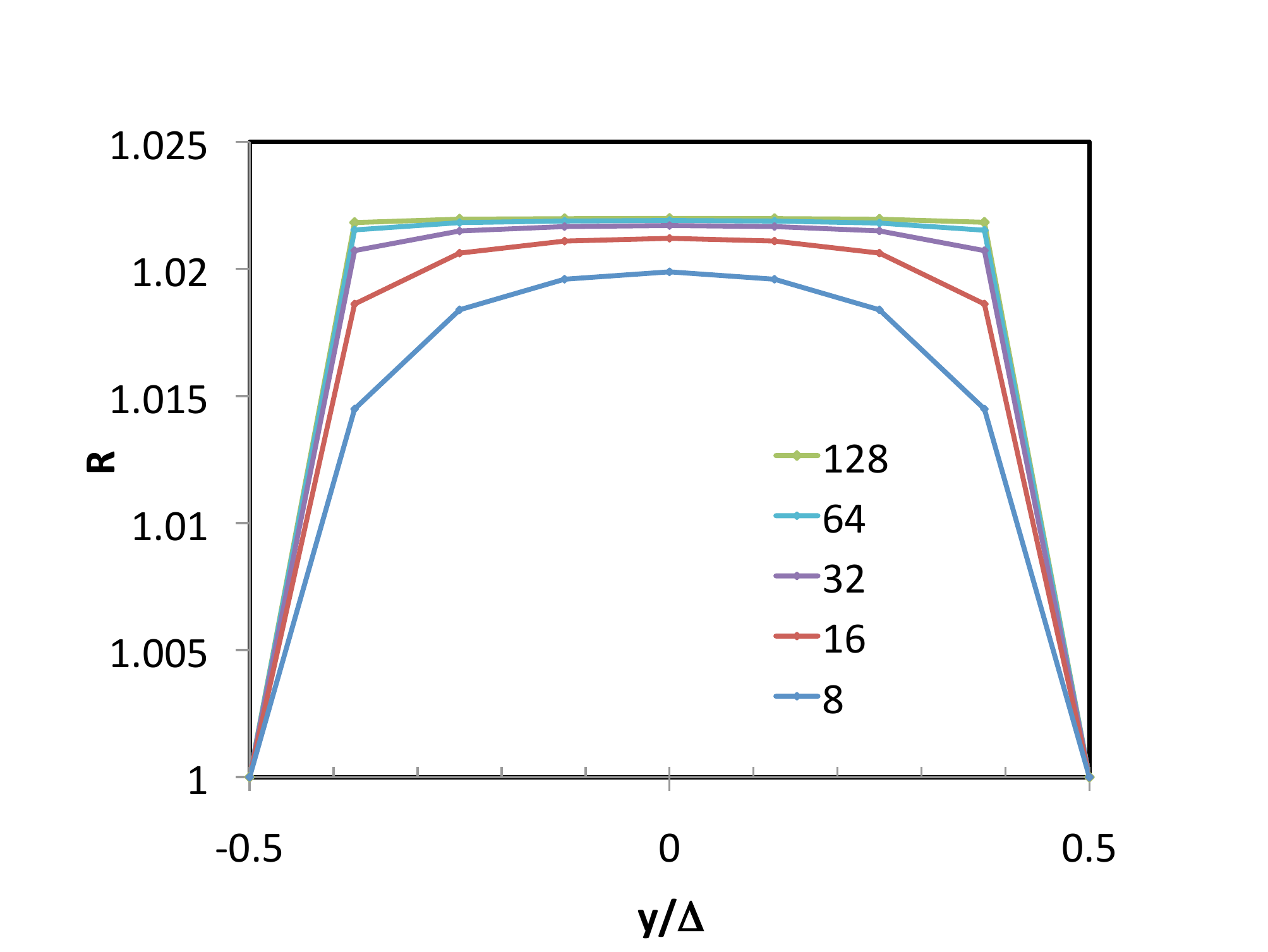}
\caption{\label{fig:Rcenter}(Color online) 
Values of $R$ along the centerline connecting the two fixed points $z_1$ and $z_2$, for various $\Delta$ (see legend) and  $L = 16384$. }
\end{figure}

In Figs.\  \ref{fig:Rclose} and \ref{fig:Rvert} we show the behavior of $R$ in both the horizontal
and vertical directions, for just the largest system $L = 16384$ and various $\Delta$,
plotted now as a function of $x/\Delta$, so that each value of the abscissa corresponds
to a similar triangle of the three points.   In general, these curves consider $x$ at much lower
values than in Fig.\ \ref{fig:Rperiodic}.  Here we see the decrease for small 
$x/\Delta$ due to the finite-size lattice effects from the points being too close together,
and the leveling out to a constant value.  The results for larger $x$ (not shown here)
exhibit roughly the same constant values for large $x$, for both the horizontal and vertical
directions.  In particular, the vertical and horizontal results both approach the same value, $1.022$.
The results along the centerline between the two points are shown in Fig.\ \ref{fig:Rcenter}.

\begin{figure}
\centering
\includegraphics[scale=0.5]{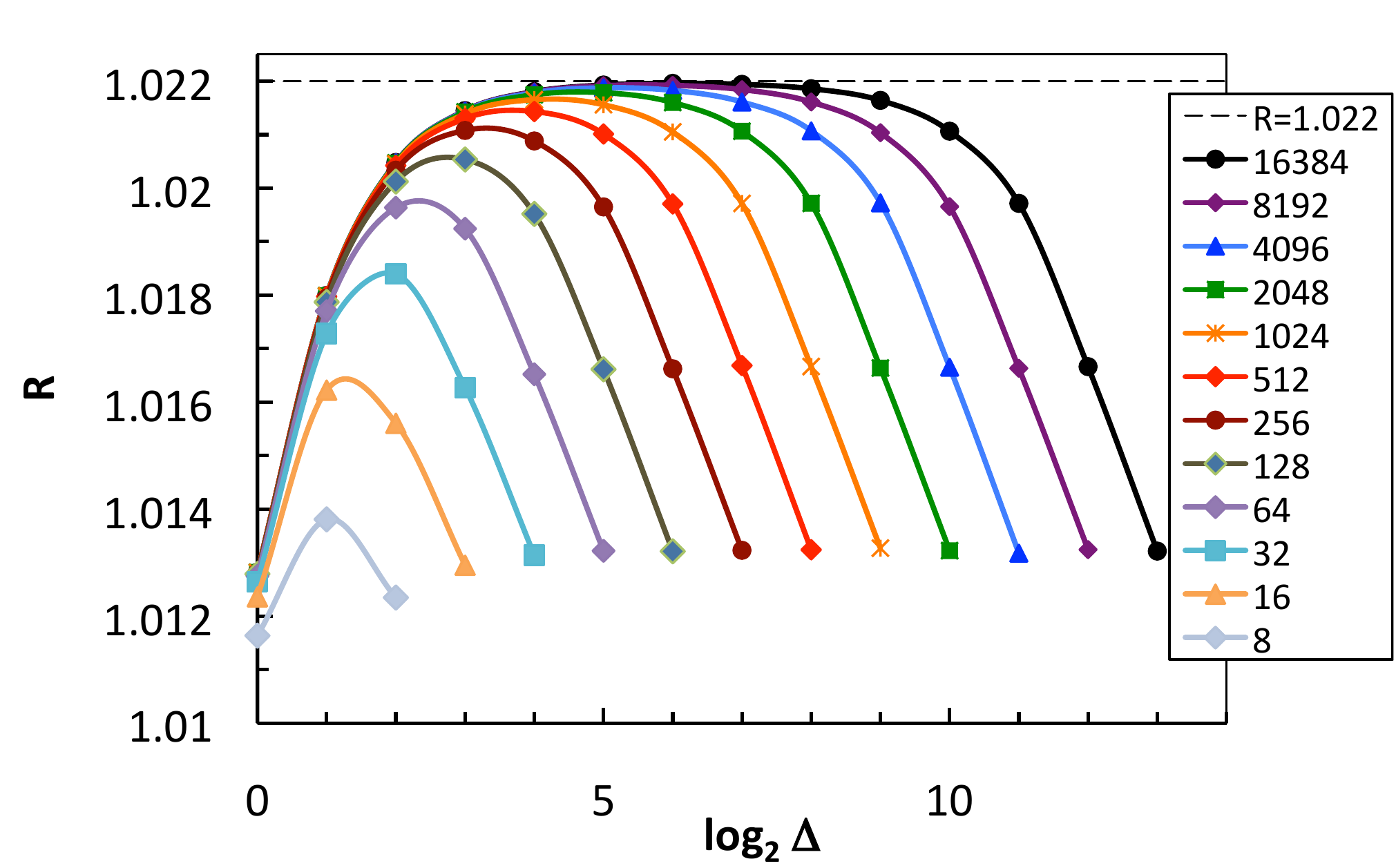}
\caption{\label{fig:Rtri}(Color online) 
Values of $R$ as a function of the side-length $\Delta$ for an equilateral triangle of points on an $L \times L$ twisted torus, with $L$ given in the legend, simulated on a triangular lattice at its bond percolation threshold $p_c = 2 \sin \pi/18$.  The errors are generally smaller than the size of the symbols.  Smoothed curves connecting the data points are drawn for ease of viewing.}
\end{figure}

\begin{figure}
\centering
\includegraphics[scale=0.5]{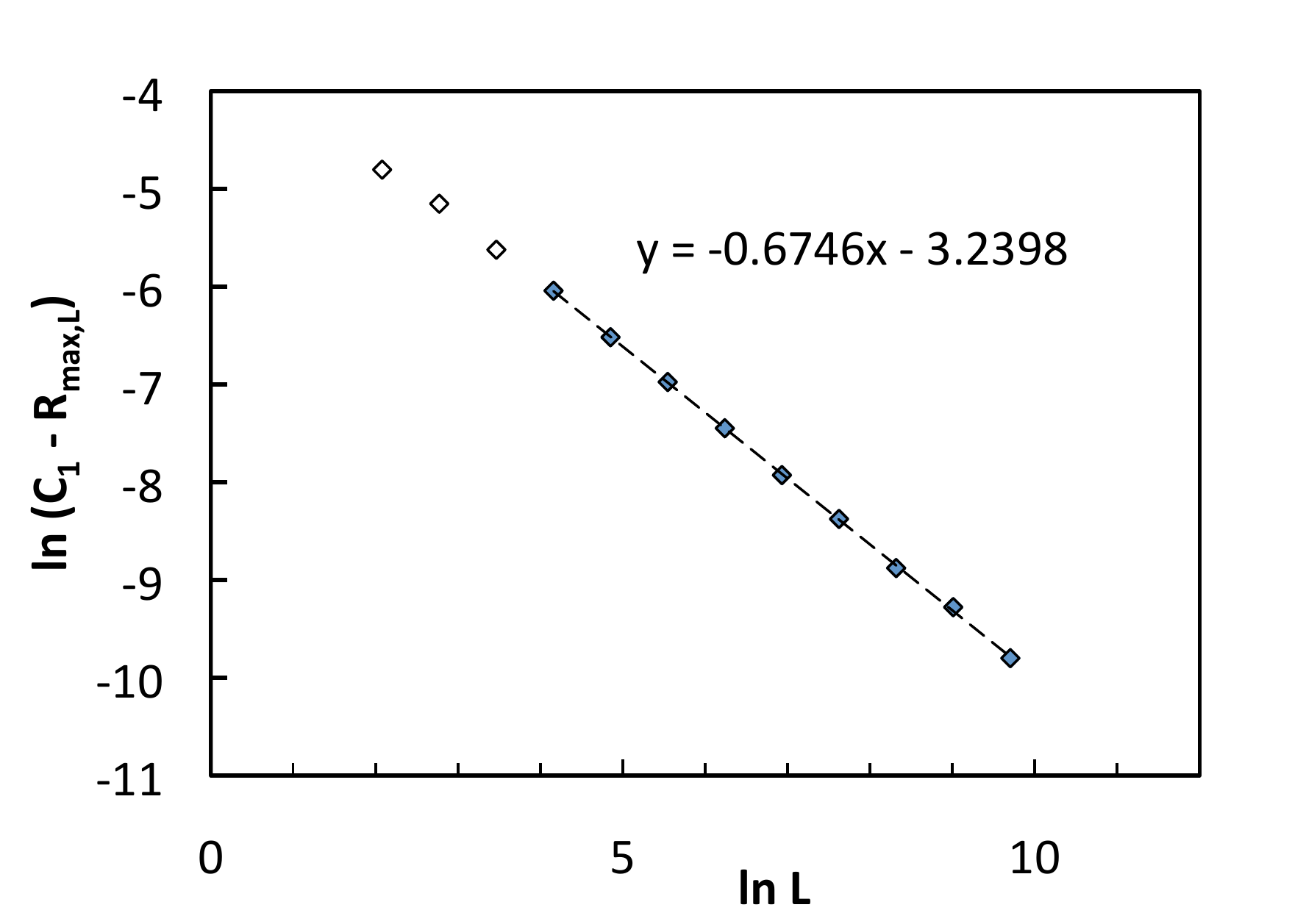}
\caption{\label{fig:Rtrimax} 
Plot of $\ln (C_1-R_\mathrm{max,L})$ vs.\ $\ln L$ for the maxima of the curves of Fig.\ \ref{fig:Rtri} using the theoretical value of  $C_1 \approx 1.022013$.  The linear fit to the
shaded points is shown in the plot, with $x$ representing $\ln L$ and $y$ representing $\ln(C_1 - R_\mathrm{max,L})$.}
\end{figure}

\begin{figure}
\centering
\includegraphics[scale=0.5]{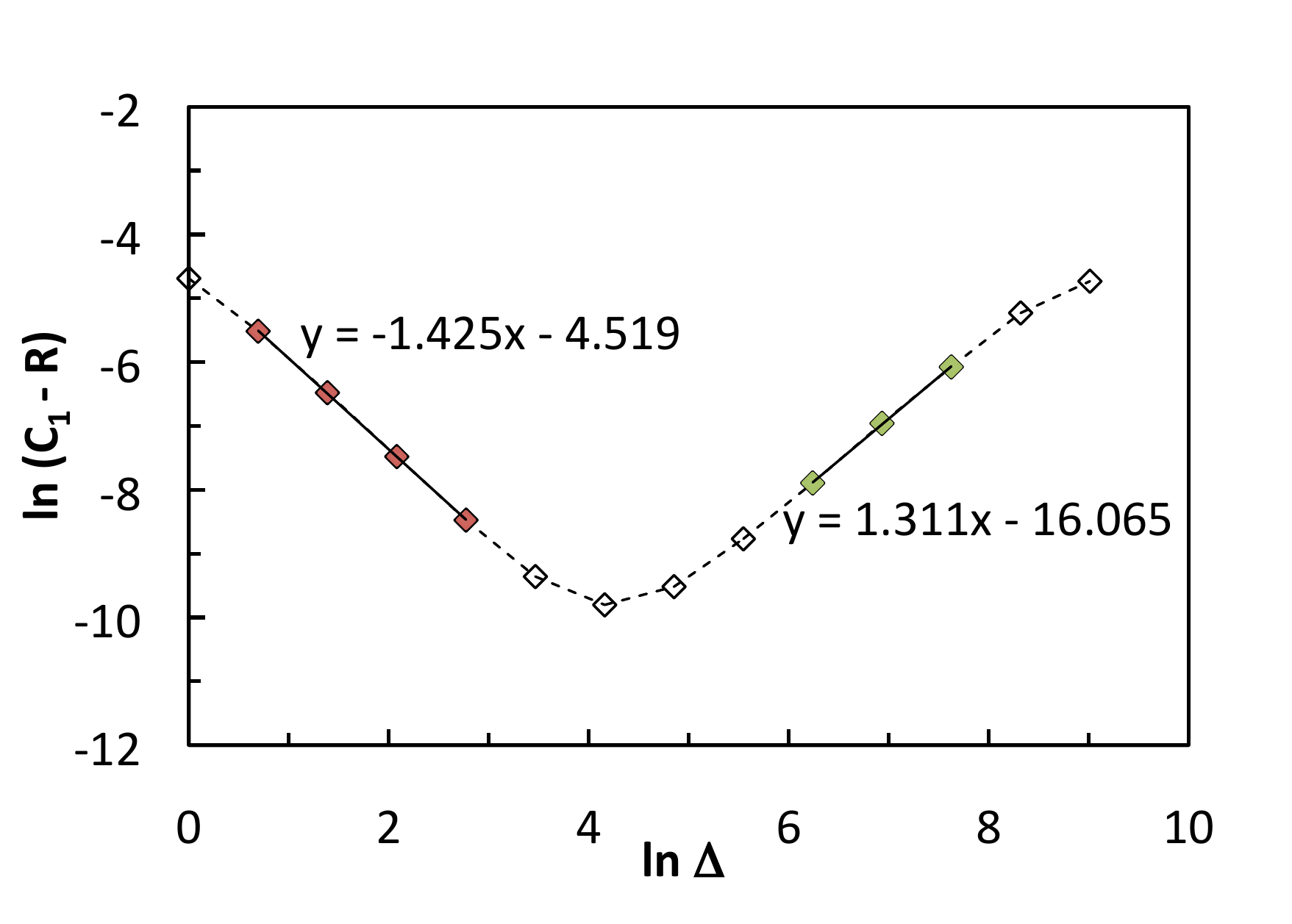}
\caption{\label{fig:RtriClose}(Color online) 
The data of Fig.\ \ref{fig:Rtri} for $L = 16384$, plotted as $\ln(C_1 - R)$ vs.\ $\ln \Delta$, using the theoretical value of $C_1 \approx 1.022013$.  The equations give the linear fits for the points $\Delta = 2$, 4, 8, and 16 (left) and $\Delta = 512, 1024,$ and $2048$ (right).
In the linear formulas, $x$ represents $\ln \Delta$ and $y$ represents $\ln(C_1 - R)$.}
\end{figure}

 \section{Point in equilateral triangle configurations on the torus}
We also considered having the three points configured as an equilateral triangle.  To do this, we used a triangular lattice, in which the periodic b.c.\ were applied on an $L \times L$ square-lattice representation with diagonal bonds, which has the effect of creating a torus with a half-twisted boundary.   Fig.\ \ref{fig:Rtri} shows $R$ as a function of the separation distance $\Delta = 1, 2, 4, \ldots L/2$ for $L = 8\ldots 16384$.   At $\Delta = L/2$, the three vertices of the triangle are equally spaced around the torus such that each pair is connected by paths of the same distance in two directions.  Flattened out and repeated, the points form a kagom\'e lattice.  For this system, we find the following behavior:
\begin{itemize}
\item As $\Delta \to 0$, $R$ decreases towards $1$ as expected, although the value even at $\Delta = 1$ (one lattice spacing apart) remains at $\approx 1.0125$.  
\item For intermediate values of $\Delta$ (of the order ${\cal O}(\sqrt{L})$), $R$ approaches $1.022$, providing further evidence for this value of $R$ for points on an infinite plane.  For $L = 16384$, the value of $R$ at the maximum is $1.02196$, just $0.00005$ below the theoretical value.   This maximum corresponds to an equilateral triangle with $\Delta = 64$.  A plot of $\ln (C_1 - R_\mathrm{max,L})$ vs.\ $\ln L$ yields a very good linear fit for $L \ge 64$ (see Fig.\ \ref{fig:Rtrimax}) implying $R_\mathrm{max,L} = C_1 - 0.0391 L^{-0.674}$.
\item As $\Delta \to L/2$, $R$ again decreases to a value of $\approx 1.0132$, which is substantially less than the maximum value $1.022$  that is found when at least two of the points are close together.  For smaller $L$, $R(\Delta = L/2)$  converges to $1.01323$ approximately as $L^{-1.5}$.  For $\Delta > L/2$, $R$ again increases due to the wraparound, so $\Delta = L/2$ is evidently a minimum point for $R$.
\end{itemize}
To study the approach to $C_1 \approx 1.022$, in Fig.\ \ref{fig:RtriClose} we plot $\ln(C_1 - R)$ vs.\ $\ln \Delta$ using the theoretical value of $C_1$.  For small $x$ we expect a power-law, and fitting the behavior in the linear regime we find a slope of about $-1.43$.  Note that in \cite{SimmonsZiffKleban09}, we found numerically for the cylinder that $C$ behaves as $\exp(-2 \pi x/L)$ where $L$ is the circumference and $x$ the distance to the end.  Transforming to the annulus this implies a decay of $R$ with two points separated by $\Delta$ (and the third far away) as $\Delta^{-1}$.  
Here we find that when all three points are separated by $\Delta$, the decay behaves as  $\approx \Delta^{-1.43}$.

For large $\Delta$, we again seem to find that $R$ behaves as a power law, here decaying with
exponent $\approx 1.3$.   All the curves for large $L$ show a similar behavior.   We have no 
explanation of why $R$ drops to this lower value as $\Delta \to L/2$.

While the curves in Fig.\ \ref{fig:Rtri} appear to be nearly symmetric, 
this symmetry is in fact an artifact of the particular system used (bond percolation).  We also considered site percolation on the
triangular lattice, where $p_c = 1/2$.  For site percolation, Eq.\ (\ref{eq:R}) must be modified
by dividing by $\sqrt{p_c}$ to account for factors of $p_c$ in the probabilities $P_2$ and $P_3$ so that
they represent conditional probabilities that the sites are occupied, and this insures that $R \to 1$ when $z_1 = z_2 = z_3$.
For large $\Delta$, the behavior is identical to that of bond percolation as seen in Fig.\ \ref{fig:Rtri},
but for small $\Delta$, the behavior is much different: while $R$ is exactly $1$ at $\Delta = 1$ (nearest
neighboring occupied sites always connect in site percolation), at $\Delta = 2$ it jumps to $\approx1.0243$ and then
drops monotonically as $\Delta$ increases, leveling at $R \approx 1.022$ in the intermediate range  $\Delta = {\cal O}(\sqrt{L})$.

 \section{Conclusions}
  
We have shown that the behavior of $R$ on a plane can be effectively studied in 
simulations on  tori of very large size, by keeping the three points  far enough apart
 to minimize finite-size effects, but also keeping the separations
of at least one pair of the points much smaller than the system size $L$.
We have confirmed the result of Delfino and Viti \cite{DelfinoViti10}
that $R$ goes to the value $1.0220...$, the same as found on a cylinder
far from the two endpoints \cite{SimmonsZiffKleban09}.
We verified this value moving $z_3$ in both the vertical and horizontal directions.
This can be seen in Figs.\ \ref{fig:Rclose} and \ref{fig:Rvert}  for larger $\Delta$ ($\ge 64$) and $x/L$ or $y/L$ greater than $2$. 
We also verified it for $z_3$ along the centerline between $z_1$ and $z_1$
(when all three points are well separated) as seen in Fig\ \ref{fig:Rcenter}.

We found the interesting result that $R \approx 1.022$ also when two of the points are very close together (though far apart compared to the 
lattice spacing), and the third anywhere on the torus.  
This behavior is consistent with conformal field theory, because in this limit $R$ only depends on the OPE coefficient, which is the same on the torus and on the plane.

We also considered the three points in an equilateral triangle configuration, on an effectively twisted torus.
For intermediate separations, $R$ goes to $1.022$, but when the three points
are far apart, $R$ drops to $1.0132$, which is the lowest value of $R$ that we have found (other than for $z_1 \to  z_2$ where  $R \to 1$).  
We have verified this behavior on the triangular lattice using both site and bond percolation.

\section{Acknowledgments}
This work was supported in part by the National Science Foundation:\ Grants Nos.\ DMS-0553487 (RMZ),
DMR-0536927 (PK) and MRSEC Grant No.\ DMR-0820054 (JJHS).

\section{Appendix.  Evaluation of $C_1$}

Here we give an expression for the constant $C_1$, which takes the value $C_1 \approx 1.022$ for percolation, that follows from
the work of Delfino and Viti \cite{DelfinoViti10} and Zamolodchikov \cite{AlZamolodchikov2005}.

Specializing Zamolodchikov's result for the three-point OPE coefficient, Eq.\ (49) of Ref.\ \cite{AlZamolodchikov2005}
for  $\alpha_1 = \alpha_2 = \alpha_3 = 1/(4 \beta) - \beta/2$, where $\beta = \sqrt{4/\kappa}$,
 with $\kappa$ the SLE parameter, and including a multiplicative factor of $\sqrt{2}$, Delfino and Viti find
 \begin{equation} \label{A1}
C_1 = 
   \frac{ \beta ^{\beta ^{-2}-\beta ^2-1}  \sqrt{2 \gamma(\beta ^2) \gamma(\beta ^{-2}-1)} \ \Upsilon_\beta \left(\frac{\beta }{2}-\frac{1}{4 \beta }\right) \Upsilon_\beta
   \left(\frac{\beta }{2}+\frac{1}{4 \beta }\right)^3}{\Upsilon_\beta(\beta )\  \Upsilon_\beta \left(\frac{1}{2 \beta }\right)^{3/2} 
   \ \Upsilon_\beta\left(\beta -\frac{1}{2 \beta }\right)^{3/2}}
\end{equation}
where $\gamma(x) := \Gamma(x)/\Gamma(1-x)$ and 
\begin{equation}
 \Upsilon_\beta(x) := \exp \int_0^{\infty } \frac{dt}{t} \left\{ 
   \left(\frac{Q}{2}-x\right)^2 e^{-t} 
   -\frac{\sinh ^2\left[\left(\frac{Q }{2}-x\right)\frac{t}{2} \right]}
   { \sinh \frac{  \beta t }{2} \sinh \frac{t}{2 \beta} }
\right\}\; 
\end{equation}
with $Q:=\beta+\beta^{-1}$.  
Using the following identities 
\begin{eqnarray}
& &\Upsilon_{\beta} (x)=\Upsilon_{1/\beta} (x)=\Upsilon_{\beta} (Q - x) \cr
& &\Upsilon_{\beta} (x+\beta )=\beta ^{1-2 \beta x } \gamma (\beta x ) \Upsilon_{\beta} (x)\cr
& &\Upsilon_{\beta} \left(Q/2 \right)=1 \cr
& &\Upsilon_{\beta} (x+1/\beta)=\beta ^{2 x/\beta-1} \gamma (x/\beta)  \Upsilon_{\beta} (x) 
\end{eqnarray}
 we can reduce (\ref{A1}) to a single integral expression
\begin{equation}
C_1 = 
\beta ^{1/2-\beta ^2}
\sqrt{ \frac{2 \gamma(\beta ^2) \gamma(\frac1 2+\frac1 4 \beta^{-2})^2} { \gamma(2-\beta ^{-2})\gamma(\frac1 2\beta ^{-2})^3}} \ \exp({I_1})
   \label{eq:C1}
   \end{equation}
where
\begin{equation}
I_1 = \int_0^{\infty } \frac{dt}{t} \left\{ 
   \frac{(1 - 2 \beta^2) e^{-t} -1}2
   -\frac{ -4 \cosh \frac{t}{4 \beta} +3 \cosh \frac{\beta t}{2}   + \cosh \frac{\beta t}{2} \cosh \frac{t}{2 \beta }}
   { 2 \sinh \frac{  \beta t }{2} \sinh \frac{t}{2 \beta} }
   \right\} 
   \label{eq:integral}
\end{equation}

Table \ref{table1} shows $C_1$ for various values of the Potts model parameter
$q$, with $\kappa = 4 \pi / [\pi - \cos^{-1} (\sqrt{q}/2)]$ in the low-density (FK-cluster) phase, 
and $\kappa' = 16/\kappa$ for the high-density (spin cluster) phase.  These values were found numerically using the Mathematica function {\tt NIntegrate[]}, increasing
the working precision to 25 digits and higher to verify the 16 digits shown here.  These values agree with those given 
in \cite{DelfinoViti10} which were quoted to just four truncated digits past the decimal point. 
Note that for $\beta = 1$ ($\kappa = 4$), the coefficient of (\ref{eq:C1}) is undefined, but taking the limit $\beta \to 1$, it converges to $\sqrt{2}\, \Gamma(3/4)/\Gamma(1/4) = 0.477988\ldots$.
For $\beta = \sqrt{3/2}$ ($\kappa = 8/3$), we can rearrange (\ref{eq:C1}) using the $\Upsilon_{\beta}$ identities to show that $C_1 = \sqrt{2}$.  This corresponds to Zamolodchikov's coefficient being exactly $1$, which is natural because $h_{1/2,0}=0$ for $\kappa=8/3$ so that the LG analysis does not distinguish between $\phi_{1/2,0}$ and the identity operator.  Using this exact result to test the accuracy of our integral expression, we indeed find $C_1 = \sqrt{2}$ to all digits of the working precision of the {\tt NIntegrate[]} function.
   
    \begin{table}[htbp]
  \centering
  \begin{tabular}{@{} |c|c|c|l| @{}}
    \hline
   $q$ & $\kappa$ &  $\beta$ & $C_1$ \\ 
  \hline
   $1$ & $6$& $\sqrt{2/3}$ &      $1.0220131331461556$ \\ 
   $2$ & $16/3$& $\sqrt{3/4}$ & $1.0524474717449139$ \\ 
   $3$ & $24/5$& $\sqrt{5/6}$ & $1.0923552364945137$ \\ 
   $4$ & $4$& $1$ &                  $1.1892071150027211 $\\
   $3$ & $10/3$& $\sqrt{6/5}$ & $1.3107927060993472 $ \\ 
      $2$ & $3$& $\sqrt{4/3}$ &  $1.3767325887917331$ \\ 
           --- & $8/3$& $\sqrt{3/2}$ & $\sqrt{2} $\\
    \hline
  \end{tabular}
  \caption{Values of $C_1$ for the $q$-state Potts model in the low-density ($\beta < 1$) and high-density ($\beta > 1$) phases; $q=1$ corresponds to percolation.}
  \label{table1}
\end{table}

 \section*{References}
\bibliographystyle{unsrt.bst}
\bibliography{ZiffSimmonsKleban10finalcorrections}

\begin{thebibliography}{10}

\bibitem{KlebanSimmonsZiff06}
P~Kleban, J~J~H Simmons, and R~M Ziff.
\newblock {Anchored critical percolation clusters and 2D electrostatics}.
\newblock {\em Phys. Rev. Lett.}, 97(11):115702, 2006.

\bibitem{SimmonsKlebanZiff07b}
J~J~H Simmons, P~Kleban, and R~M Ziff.
\newblock {Exact factorization of correlation functions in two-dimensional
  critical percolation}.
\newblock {\em Phys. Rev. E}, 76(4):41106, 2007.

\bibitem{SimmonsZiffKleban09}
J~J~H Simmons, R~M Ziff, and P~Kleban.
\newblock Factorization of percolation density correlation functions for
  clusters touching the sides of a rectangle.
\newblock {\em J. Stat. Mech. Th. Exp.}, 2009(02):P02067, 2009.

\bibitem{HonglerSmirnov10}
C~Hongler and S~Smirnov.
\newblock Critical percolation: the expected number of clusters in a rectangle.
\newblock {\em Probability Theory and Related Fields (online first)}, pages
  1--22, 2010.

\bibitem{KarsaiKovacsAnglesdAuriacIgloi08}
M~Karsai, I~A Kov\'acs, J-Ch Angl\`es~d'Auriac, and F~Igl\'oi.
\newblock Density of critical clusters in strips of strongly disordered
  systems.
\newblock {\em Phys. Rev. E}, 78(6):061109, 2008.

\bibitem{Ridout10}
D~Ridout.
\newblock {$(sl)_{-1/2}$ and the triplet model}.
\newblock {\em Nucl. Phys. B}, 835:314--342, 2010.

\bibitem{MathieuRidout07}
P~Mathieu and D~Ridout.
\newblock {From percolation to logarithmic conformal field theory}.
\newblock {\em Phys. Lett. B}, 657(1-3):120--129, 2007.

\bibitem{MathieuRidout08}
P~Mathieu and D~Ridout.
\newblock {Logarithmic $M(2,p)$ minimal models, their logarithmic couplings,
  and duality}.
\newblock {\em Nucl. Phys. B}, 801(3):268--295, 2008.

\bibitem{Ridout09}
D~Ridout.
\newblock {On the percolation BCFT and the crossing probability of Watts}.
\newblock {\em Nucl. Phys. B}, 810(3):503--526, 2009.

\bibitem{SimmonsCardy09}
J~J~H Simmons and J~Cardy.
\newblock Twist operator correlation functions in {O}$(n)$ loop models.
\newblock {\em J. Phys. A}, 42(23):235001, 2009.

\bibitem{SheffieldWilson10}
S~Sheffield and D~B Wilson.
\newblock {Schramm's proof of Watts' formula}.
\newblock {\em arXiv preprint arXiv:1003.3271}.

\bibitem{DelfinoViti10}
G~Delfino and J~Viti.
\newblock On three-point connectivity in two-dimensional percolation.
\newblock {\em J. Phys. A}, 44:032001, 2011.

\bibitem{Polyakov70}
A~M Polyakov.
\newblock Conformal symmetry of critical fluctuations.
\newblock {\em JETP Lett.}, 12(12):381--383, 1970.

\bibitem{Dotsenko85}
Vl~S Dotsenko and V~A Fateev.
\newblock Operator algebra of two-dimensional conformal theories with central
  charge $c <= 1$.
\newblock {\em Phys. Lett. B}, 154(4):291--295, 1985.

\bibitem{AlZamolodchikov2005}
Al~B Zamolodchikov.
\newblock Three-point function in the minimal {L}iouville gravity.
\newblock {\em Theor. and Math. Phys.}, 142:183--196, 2005.

\end{thebibliography}

\end{document}